# Fluctuation Electromagnetic Interaction Between Rotating Nanoparticle and Near-Field of the Surface

A.A. Kyasov, G.V. Dedkov

Nanoscale Physics Group, Kabardino-Balkarian State University, Nalchik, Russia

Fluctuation electromagnetic interaction between a small rotating particle and a polarizable surface produces the friction torque on the particle, changes the attraction force (van -der -Waals force) and the rate of heat exchange in this system. In this paper we consider the case where the rotation axis is parallel to the surface. We present general theoretical basis allowing to obtain characteristics of the interaction at an arbitrary orientation of the rotation axis. In comparison to the case where the rotation axis is perpendicular to the surface (arXiv: 1208.2603), the obtained integral expressions for the friction torque, interaction energy and heating rate have different numerical coefficients. Numerical estimations are given in the case Au (particle)- Au(surface) and SiC (particle)-SiC (surface).

## 1. Introduction

In the preceding papers [1,2] we have calculated the force of attraction, friction torque and heating of a rotating nanoparticle in the near-field of the surface in particular case where the axis of particle rotation is perpendicular to the surface. In this work we consider the case where the rotation axis is parallel to the surface. We present a detailed theoretical basis allowing to calculate the corresponding quantities in the case of arbitrary orientation of the rotation axis relative to the surface. Previous results [1,2] can be easily retrieved when turning the rotation axis. Comparing the particle-surface friction torque with that corresponding to the particle rotation in the vacuum background [3], we conclude that in our case the friction turns out to be higher by 5 to 10 orders of magnitude. This is of paramount importance in development of NEMS where the corresponding interactions must be taken into account.

## 2. Theory

Figure 1 shows the considered system, which consists of an isotropic particle at temperature $T_1$ placed in vacuum at a distance $z_0$ from a polarizable surface at temperature $T_2$. The particle rotates around its $x$ axis with the angular frequency $\Omega$ assuming to be a point-like fluctuating nonrelativistic dipole. In this case the following conditions are fulfilled ($\omega_0$ is the characteristic absorption frequency of the particle, $c, k_B,$ and $\hbar$ are the speed of light in vacuum, Boltzmann's and Planck's constants)

$$R \ll \min\left\{\frac{2\pi c}{\omega_0}, \frac{2\pi c}{\Omega}, \frac{2\pi \hbar c}{k_B T_1}, \frac{2\pi \hbar c}{k_B T_2}\right\}, R \ll z_0$$

## 2.1 Calculation of the frictional torque

The torque $\mathbf{M} = (M_x, 0, 0)$ is given by

$$M_x = \langle [\mathbf{dE}]_x \rangle = \langle [\mathbf{d}^{sp}\mathbf{E}^{in}]_x \rangle + \langle [\mathbf{d}^{in}\mathbf{E}^{sp}]_x \rangle = M_x^{(1)} + M_x^{(2)}$$

where $\langle ... \rangle$ denotes total quantum-statistical averaging, the superscripts "$sp$" and "$in$" denote the spontaneous and induced quantities taken in the resting reference frame $\Sigma$ $(x, y, z)$ related to the surface. The contribution $M_x^{(1)}$ that is associated with the spontaneous dipole moment of the particle $\mathbf{d}^{sp}$ is given by

$$M_x^{(1)} = \langle [\mathbf{d}^{sp}\mathbf{E}^{in}]_x \rangle = \langle d_y^{sp} E_z^{in} - d_z^{sp} E_y^{in} \rangle \tag{1}$$

The values of $\mathbf{d}^{sp}(t)$ and $\mathbf{E}^{in}(\mathbf{r},t)$ are represented by the Fourier transforms

$$\mathbf{d}^{sp}(t) = \int_{-\infty}^{+\infty} \frac{d\omega}{2\pi} \mathbf{d}^{sp}(\omega) \exp(-i\omega t) \tag{2}$$

$$\mathbf{E}^{in}(x, y, t) = \int \frac{d\omega}{2\pi} \frac{d^2 k}{(2\pi)^2} \mathbf{E}^{in}(\omega \mathbf{k}, z) \exp(i(k_x x + k_y y - \omega t)) \tag{3}$$

Substituting (2) and (3) (taken at the particle location point $\mathbf{r}_0 = (0, 0, z_0)$) in (1) yields

$$M_x^{(1)} = \int \frac{d\omega'}{2\pi} \frac{d\omega}{2\pi} \frac{d^2 k}{(2\pi)^2} \exp(-i(\omega + \omega')t) \langle d_y^{sp}(\omega') E_z^{in}(\omega \mathbf{k}, z_0) - d_z^{sp}(\omega') E_y^{in}(\omega \mathbf{k}, z_0) \rangle \tag{4}$$

The components $\mathbf{E}^{in}(\omega \mathbf{k}, z)$ are expressed through the gradient of the Fourier transform of the induced potential of the surface

$$\phi^{in}(\omega,z) = \frac{2\pi}{k}\Delta(\omega)\exp(-k(z+z_0))\left[ik_x d_x^{sp}(\omega) + ik_y d_y^{sp}(\omega) + kd_z^{sp}(\omega)\right] \quad (5)$$

$$E_x^{in}(\omega\mathbf{k},z) = -ik_x\phi^{in}(\omega\mathbf{k},z),\ E_y^{in}(\omega\mathbf{k},z) = -ik_y\phi^{in}(\omega\mathbf{k},z),\ E_z^{in}(\omega\mathbf{k},z) = k\phi^{in}(\omega\mathbf{k},z) \quad (6)$$

$$\Delta(\omega) = \frac{\varepsilon(\omega)-1}{\varepsilon(\omega)+1} \quad (7)$$

where $\varepsilon(\omega)$ is the dielectric permittivity of the surface and $k = |\mathbf{k}| = (k_x^2 + k_y^2)^{1/2}$. From (4)—(6) we obtain

$$M_x^{(1)} = \frac{1}{(2\pi)^4}\int d\omega' d\omega d^2k \exp(-(\omega+\omega')t)(2\pi/k)\Delta(\omega)\exp(-2kz_0)\cdot$$
$$\cdot\left\{\begin{array}{l}\langle d_y^{sp}(\omega')k(ik_x d_x^{sp}(\omega) + ik_y d_y^{sp}(\omega) + kd_z^{sp}(\omega))\rangle \\ -\langle d_z^{sp}(\omega')(-ik_y)(ik_x d_x^{sp}(\omega) + ik_y d_y^{sp}(\omega) + kd_z^{sp}(\omega))\rangle\end{array}\right\} \quad (8)$$

It should be noted once again that all components of the Fourier-transform of the spontaneous dipole moment of the particle in Eq. (8) are taken in the resting reference system $\Sigma\,(x,y,z)$. For calculating correlators in Eq. (8), however, we have to perform the transformation of the corresponding Fourier components from the system $\Sigma\,(x,y,z)$ to the resting frame of the particle $\Sigma'\,(x',y',z')$. This is done using the following relations (see Appendix A for more details)

$$d_x^{sp}(\omega) = d_x^{sp'}(\omega)$$
$$d_y^{sp}(\omega) = \frac{1}{2}\left(d^{sp'}_y(\omega^+) + d^{sp'}_y(\omega^-) + id^{sp'}_z(\omega^+) - id^{sp'}_z(\omega^-)\right)$$
$$d_z^{sp}(\omega) = \frac{1}{2}\left(-id^{sp'}_y(\omega^+) + id^{sp'}_y(\omega^-) + d^{sp'}_z(\omega^+) + d^{sp'}_z(\omega^-)\right) \quad (9)$$

where $\omega^\pm = \omega\pm\Omega$.

The fluctuation-dissipation relationships are given in the particle rest frame $\Sigma'\,(x',y',z')$ and have a usual form

$$\langle d_i^{sp'}(\omega')d_k^{sp'}(\omega)\rangle = 2\pi\delta_{ik}\delta(\omega+\omega')\hbar\alpha''(\omega)\coth\frac{\hbar\omega}{2k_B T_1} \quad (10)$$

After from (8)—(10) after simple algebra we obtain

$$M_x^{(1)} = -\frac{3\hbar}{16\pi z_0^3} \int_0^\infty d\omega \Delta''(\omega) \left[ \alpha''(\omega^-) \coth \frac{\hbar\omega^-}{2k_B T_1} - \alpha''(\omega^+) \coth \frac{\hbar\omega^+}{2k_B T_1} \right] \quad (11)$$

The contribution $M_x^{(2)}$ that is associated with the spontaneous fluctuating field of the surface $\mathbf{E}^{sp}$ is given by

$$M_x^{(2)} = \left\langle [\mathbf{d}^{in} \mathbf{E}^{sp}]_x \right\rangle = \left\langle d_y^{in} E_z^{sp} - d_z^{in} E_y^{sp} \right\rangle \quad (12)$$

Inserting the Fourier frequency-expansions of $\mathbf{d}^{in}(t)$ and $\mathbf{E}^{sp}(\mathbf{r}_0, t)$ in (12) yields

$$M_x^{(2)} = \int \frac{d\omega'}{2\pi} \frac{d\omega}{2\pi} \exp(-i(\omega+\omega')t) \left\langle d_y^{in}(\omega) E_z^{sp}(\mathbf{r}_0, \omega') - d_z^{in}(\omega) E_y^{sp}(\mathbf{r}_0, \omega') \right\rangle \quad (13)$$

Relation between the Fourier transforms of $\mathbf{d}^{in}(t)$ for rotating particle and the field $\mathbf{E}^{sp}(\mathbf{r}_0, t)$ of the surface in the reference frame $\Sigma\,(x, y, z)$ are cast in the form (see Appendix B)

$$d_x^{in}(\omega) = \alpha(\omega) E^{sp}{}_x(\omega)$$

$$d_y^{in}(\omega) = \frac{1}{2}\left[ \alpha(\omega^+)\left( E^{sp}{}_y(\omega) + iE^{sp}{}_z(\omega) \right) + \alpha(\omega^-)\left( E^{sp}{}_y(\omega) - iE^{sp}{}_z(\omega) \right) \right]$$

$$d_z^{in}(\omega) = \frac{1}{2}\left[ -i\alpha(\omega^+)\left( E^{sp}{}_y(\omega) + iE^{sp}{}_z(\omega) \right) + i\alpha(\omega^-)\left( E^{sp}{}_y(\omega) - iE^{sp}{}_z(\omega) \right) \right] \quad (14)$$

where we have omitted argument $\mathbf{r}_0$ for brevity. Substituting (14) in (13) and making use some transformations yields

$$M_x^{(2)} = \frac{i}{2} \int \frac{d\omega'}{2\pi} \frac{d\omega}{2\pi} \exp(-i(\omega+\omega')t) \left( \alpha(\omega^+) - \alpha(\omega^-) \right) \cdot$$
$$\cdot \left[ \left\langle E_z^{sp}(\omega) E_z^{sp}(\omega') \right\rangle + \left\langle E_y^{sp}(\omega) E_y^{sp}(\omega') \right\rangle \right]. \quad (15)$$

Correlators in Eq. (15) are calculated using the fluctuation-dissipation relations, which can be conveniently represented as follows (at the point $\mathbf{r}_0 = (0, 0, z_0)$)

$$\begin{cases} \langle E_i^{sp}(\omega) E_i^{sp}(\omega') \rangle = 2\pi\delta(\omega+\omega')\left(E^{sp}_i\right)^2_\omega, \; i=x,y,z \\ \left(E^{sp}_i\right)^2_\omega = \int \frac{d^2 k}{(2\pi)^2}\left(E^{sp}_i\right)^2_{\omega \mathbf{k}} \\ \left(E^{sp}_x\right)^2_{\omega \mathbf{k}} = \hbar \coth\frac{\hbar\omega}{2k_B T_2}\Delta''(\omega)\frac{2\pi}{k}k_x^2 \exp(-2k z_0) \\ \left(E^{sp}_y\right)^2_{\omega \mathbf{k}} = \hbar \coth\frac{\hbar\omega}{2k_B T_2}\Delta''(\omega)\frac{2\pi}{k}k_y^2 \exp(-2k z_0) \\ \left(E^{sp}_z\right)^2_{\omega \mathbf{k}} = \hbar \coth\frac{\hbar\omega}{2k_B T_2}\Delta''(\omega)\frac{2\pi}{k}k^2 \exp(-2k z_0) \end{cases} \qquad (16)$$

It should be noted that no summation over "" in Eq. (16)

From (15), (16) we obtain

$$M_x^{(2)} = -\frac{3\hbar}{16\pi z_0^3}\int_0^\infty d\omega \Delta''(\omega)\left[\alpha''(\omega^+)\frac{\hbar\omega}{2k_B T_2} - \alpha''(\omega^-)\frac{\hbar\omega}{2k_B T_2}\right] \qquad (17)$$

Finally, summing (11) and (17) yields

$$M_x = -\frac{3\hbar}{16\pi z_0^3}\int_0^\infty d\omega \Delta''(\omega)\cdot \begin{cases} \alpha''(\omega^-)\left[\coth\frac{\hbar\omega^-}{2k_B T_1} - \coth\frac{\hbar\omega}{2k_B T_2}\right] - \\ -\alpha''(\omega^+)\left[\coth\frac{\hbar\omega^+}{2k_B T_1} - \coth\frac{\hbar\omega}{2k_B T_2}\right] \end{cases} \qquad (18)$$

Coefficient 3/4 in Eq. (18) differs from that obtained in [1,2] where the rotation axis of the particle is perpendicular to the surface. As well, formula (18) can be written in a more compact form

$$M_x = \frac{3\hbar}{16\pi z_0^3}\int_{-\infty}^{+\infty} d\omega \Delta''(\omega)\alpha''(\omega^+)\left[\coth\frac{\hbar\omega^+}{2k_B T_1} - \coth\frac{\hbar\omega}{2k_B T_2}\right] \qquad (19)$$

## 2.2 Calculation of the free energy (interaction potential) and heating rate

Other quantities that characterize fluctuation-electromagnetic interaction in the system particle-surface, such as free energy (interaction potential) $U$ and heating (cooling) rate $\dot{Q}$ are calculated quite analogously. The starting equations have the form

$$U = -\frac{1}{2}\langle \mathbf{dE} \rangle = -\frac{1}{2}\langle \mathbf{d}^{sp}\mathbf{E}^{in}\rangle - \frac{1}{2}\langle \mathbf{d}^{in}\mathbf{E}^{sp}\rangle \tag{20}$$

$$\dot{Q} = \langle \mathbf{\dot{d}E}\rangle = \langle \mathbf{\dot{d}}^{sp}\mathbf{E}^{in}\rangle + \langle \mathbf{\dot{d}}^{in}\mathbf{E}^{sp}\rangle \tag{21}$$

When calculating the contributions associated with $\mathbf{d}^{sp}$ (the first terms in (20), (21)) one must use Eqs. (9),(10), while in calculating the contributions associated with $\mathbf{E}^{sp}$ (the second terms in (20), (21))---Eqs. (14) and (16).

As a result we obtain

$$U = -\frac{\hbar}{32\pi z_0^3}\int_0^{+\infty} d\omega \left\{ \begin{array}{l} 2\left[\Delta'(\omega)\alpha''(\omega)\coth\frac{\hbar\omega}{2k_BT_1} + \Delta''(\omega)\alpha'(\omega)\coth\frac{\hbar\omega}{2k_BT_2}\right] + \\ +3\left[\Delta'(\omega)\alpha''(\omega^-)\coth\frac{\hbar\omega^-}{2k_BT_1} + \Delta''(\omega)\alpha'(\omega^-)\coth\frac{\hbar\omega}{2k_BT_2}\right] + \\ +3\left[\Delta'(\omega)\alpha''(\omega^+)\coth\frac{\hbar\omega^+}{2k_BT_1} + \Delta''(\omega)\alpha'(\omega^+)\coth\frac{\hbar\omega}{2k_BT_2}\right] \end{array} \right\} \tag{22}$$

$$\dot{Q} = \frac{\hbar}{16\pi z_0^3}\int_0^{+\infty} d\omega\omega \left\{ \begin{array}{l} 2\Delta''(\omega)\alpha''(\omega)\left[\coth\frac{\hbar\omega}{2k_BT_2} - \coth\frac{\hbar\omega}{2k_BT_1}\right] + \\ +3\Delta''(\omega)\alpha''(\omega^-)\left[\coth\frac{\hbar\omega}{2k_BT_2} - \coth\frac{\hbar\omega^-}{2k_BT_1}\right] + \\ +3\Delta''(\omega)\alpha''(\omega^+)\left[\coth\frac{\hbar\omega}{2k_BT_2} - \coth\frac{\hbar\omega^+}{2k_BT_1}\right] \end{array} \right\} \tag{23}$$

In particular case $\Omega = 0$ formulas (22) and (23) coincide with the well-known results corresponding to the static configuration, while $M_x = 0$.

## 3. Numerical calculation of stopping times of nanoparticles

It is interesting to compare stopping times of small rolling nanoparticles in the vacuum background and near a heated surface on the one hand, and stopping times under uniform motion with the constant velocity parallel to the surface and in the vacuum background, on the other hand. In the isothermal case where the particles and the surface have the same temperature $T$, the frictional force acting on a particle moving with the constant velocity $V$ (bearing in mind that we assume the nonrelativistic case) is given by [4]

$$F_x = -\frac{3\hbar V}{8\pi z_0^5} \int_{-\infty}^{+\infty} d\omega \Delta''(\omega)\alpha''(\omega)\left(-\frac{\partial}{\partial \omega}\right)\coth\frac{\hbar\omega}{2k_B T} \qquad (24)$$

Equation (19) in this case takes the form

$$M_x = \frac{3\hbar}{16\pi z_0^3} \int_{-\infty}^{+\infty} d\omega \Delta''(\omega)\alpha''(\omega)\left(-\frac{d}{d\omega}\right)\coth\frac{\hbar\omega}{2k_B T} \qquad (25)$$

When the particle rotates or uniformly moves in free vacuum, equations similar to Eq. (25) and Eq. (24) were first obtained in [3] and [5] (in the latter case see also [6])

$$M_x^{(vac)} = -\frac{2\hbar\Omega}{3\pi c^3} \int_{-\infty}^{+\infty} d\omega\, \omega^3 \Delta''(\omega)\alpha''(\omega)\left(-\frac{\partial}{\partial\omega}\right)\coth\frac{\hbar\omega}{2k_B T} \qquad (26)$$

$$F_x = -\frac{\hbar^2 V}{3\pi c^5 k_B T} \int_0^{\infty} d\omega\, \omega^5 \alpha''(\omega)\sinh^{-2}(\hbar\omega/2k_B T) \qquad (27)$$

Using Eqs. (24)—(27) and Newton's second law, the corresponding stopping times can be written as (note that $J_{\Omega,s}(\omega_W) = J_{V,s}(\omega_W)$)

$$\begin{aligned}
\tau_{\Omega,s} &= \frac{128\pi^2}{45}\frac{\rho R^5}{\hbar} z_0^3 / J_{\Omega,s}(\omega_W), \\
\tau_{V,s} &= \frac{32\pi^2}{9}\frac{\rho R^5}{\hbar}\frac{z_0^5}{R^2} / J_{V,s}(\omega_W), \\
\tau_{\Omega,vac} &= \frac{4\pi^2}{5}\frac{\rho R^5}{\hbar} c^3 / J_{\Omega,vac}(\omega_W), \\
\tau_{V,vac} &= 4\pi^2 \frac{\rho R^3}{\hbar} c^5 \omega_W / J_{V,vac}(\omega_W)
\end{aligned} \qquad (28)$$

where the subscripts *s, vac* denote "surface" and "vacuum", $\Omega, V$ are the angular and linear velocities, $\omega_W = k_B T/\hbar$ is the Wien frequency, and the functions $J_{k,m}(\omega_W), k=\Omega, V; m=s, vac$ denote the corresponding frequency integrals in (19), (24)—(26). We also took into account that the mass and the moment inertia of spherical particle are $4\pi\rho R^5/3$ and $8\pi\rho R^5/15$, where $\rho$ and $R$ denote the density and radius of the particle. Several important conclusions can be obtained from (28) at a glance. From the first two we see that $\tau_{\Omega,s}/\tau_{V,s} = \frac{4}{5}(R/z_0)^2 \ll 1$ , since $R \ll z_0$. Moreover, the stopping times for a uniformly moving particle (both in vacuum and near a surface) do not depend on its radius $R$, while the stopping times of rotating particles are proportional to $R^2$. As follows from the structure of integrals in (28), all stopping times monotonously decrease with increasing temperature $T$. More accurate numerical estimations of the stopping times are shown in Figs. 2,3 in particular cases of a $SiC$ particle with $R = 1 nm$ above a $SiC$ surface and an $Au$ particle above a gold surface at $T = 300, 600 K$ (or in the vacuum background with the same temperature). The dielectric functions were approximated in the form

$$\varepsilon(\omega) = \varepsilon_\infty + \frac{\omega_T^2(\varepsilon_0 - \varepsilon_\infty)}{\omega_T^2 - \omega^2 - i\gamma\omega} \qquad (29)$$

$$\varepsilon(\omega) = 1 - \frac{\omega_p^2}{\omega(\omega + i/\tau)} \qquad (30)$$

and the particle polarizability was taken to be

$$\alpha(\omega) = R^3 \frac{(\varepsilon(\omega) - 1)}{(\varepsilon(\omega) + 2)} \qquad (31)$$

Parameters of $SiC$ and $Au$ correspond to [7]: $\varepsilon_0 = 9.8$, $\varepsilon_\infty = 6.7$, $\omega_T = 0.098\, eV$, $\gamma = 0.00585\, eV$ $\omega_p = 1.37 \cdot 10^{16}\, s^{-1}$, $\tau = 1.89 \cdot 10^{-14}\, s$.

As we can see from Figs. 2,3, the stopping times of rolling nanoparticles in the near-field of the surface are much shorter than in the vacuum background and in the cases of uniform motion of the particle near the surface and in free vacuum. For metallic particles, the dominating contribution in Eq. (26) results from the magnetic polarizability of the particle [3], that can be higher by $1 \div 2$ orders of magnitude than the contribution from electric polarizability. But generally, the estimation $\tau_\Omega / \tau_\Omega^{(vac)} \ll 1$ proves to be valid in this case, as well.

## 4. CONCLUSIONS

Using the fluctuation electromagnetic theory, we have obtained closed nonrelativistic expressions for the friction torque, attraction force and heating rate of a small spherical particle rotating in the near-field of the surface. Material properties of the particle and the surface are characterized by the frequency-dependent polarizability and dielectric permittivity. The temperatures of the particle and the surface are assumed to be arbitrary. It is worth noting that an anisotropy of the particle polarizability has no appreciable physical importance in this case and may only change numerical factors in the quantities under study. Apparently, it is not difficult to generalize the obtained formulas in this case.

Assuming the isothermal conditions, we have compared stopping times of spherical SiC and Au particles corresponding to the rolling and uniform motion near the surfaces of SiC and Au, and those for rolling and uniform motion near the surface and in the vacuum background. In both of the cases the stopping times for rolling in the near-field of the surface turn out to be smaller by 5 to 9 orders of magnitude than in free vacuum and under uniform motion near the surface. This is of paramount importance in development of NEMS where the corresponding dissipative interactions must be taken into account.

## 5. Mathematical Appendices

**Appendix A: Coordinate systems $\Sigma\,(x,y,z)$ and $\Sigma'\,(x',y',z')$**

Consider the coordinate systems shown in Fig. 1. System $\Sigma\,(x,y,z)$ is related to the surface that is assumed to be at rest. System $\Sigma'\,(x',y',z')$ rotates with the particle with the angular velocity $\mathbf{\Omega}=(\Omega,0,0)$ around the resting axis $x'$ which is parallel to the axis $x$ of the system $\Sigma\,(x,y,z)$. The components of any vector $\mathbf{A}=(A_x,A_y,A_z)$ in the system $\Sigma$ are related to the components of $\mathbf{A}'$ in $\Sigma'$ by

$$\begin{pmatrix} A_x(t) \\ A_y(t) \\ A_z(t) \end{pmatrix} = \begin{pmatrix} 1 & 0 & 0 \\ 0 & \cos\Omega t & -\sin\Omega t \\ 0 & \sin\Omega t & \cos\Omega t \end{pmatrix} \begin{pmatrix} A'_x(t) \\ A'_y(t) \\ A'_z(t) \end{pmatrix} \qquad (A1)$$

The inverse transformation has the form

$$\begin{pmatrix} A'_x(t) \\ A'_y(t) \\ A'_z(t) \end{pmatrix} = \begin{pmatrix} 1 & 0 & 0 \\ 0 & \cos\Omega t & \sin\Omega t \\ 0 & -\sin\Omega t & \cos\Omega t \end{pmatrix} \begin{pmatrix} A_x(t) \\ A_y(t) \\ A_z(t) \end{pmatrix} \qquad (A2)$$

In general case the components of $\mathbf{A}$ are the functions of time, as seen from (A1) and (2). Moreover, within the nonrelativistic statement of the problem the time $t$ is the same in the left and right sides of (A1), (A2). Obviously, this is true at

$$\Omega R \ll c \qquad (A3)$$

where $R$ is the particle radius, and $c$ is the speed of light.

**Appendix B: Relation between the Fourier components of spontaneous dipole moment of spherical particle in systems $\Sigma$, $\Sigma'$**

Applying Eq.(A1) to the vector $\mathbf{d}^{sp}(t)$ we obtain

$$d_x^{sp}(t) = d^{sp}{}'_x(t)$$

$$d_y^{sp}(t) = d^{sp}{}_y{}'(t)\cos\Omega t - d^{sp}{}_z{}'(t)\sin\Omega t \qquad (B1)$$

$$d_z^{sp}(t) = d^{sp}{}_y{}'(t)\sin\Omega t + d^{sp}{}_z{}'(t)\cos\Omega t$$

Making use the Fourier transformation in both sides of (A1)

$$\mathbf{d}(\omega) = \int_{-\infty}^{+\infty} dt\, \mathbf{d}(t)\exp(i\omega t),$$

we obtain

$$d_x^{sp}(\omega) = d_x^{sp\,\prime}(\omega)$$

$$d_y^{sp}(\omega) = \frac{1}{2}\left(d^{sp}{}_y{}'(\omega^+) + d^{sp}{}_y{}'(\omega^-) + id^{sp}{}_z{}'(\omega^+) - id^{sp}{}_z{}'(\omega^-)\right) \qquad (B2)$$

$$d_z^{sp}(\omega) = \frac{1}{2}\left(-id^{sp}{}_y{}'(\omega^+) + id^{sp}{}_y{}'(\omega^-) + d^{sp}{}_z{}'(\omega^+) + d^{sp}{}_z{}'(\omega^-)\right)$$

where $\omega^\pm = \omega \pm \Omega$.

Equations (B2) allows one to use the fluctuation-dissipation relation for the spontaneous dipole moment of the particle in its resting frame $\Sigma'$.

**Appendix C: Relation between the Fourier transforms of the induced dipole moment of a rolling particle and the spontaneous electric field of the surface in the system $\Sigma$**

Applying the same procedure (as in Appendix B) to the vector of the induced dipole moment of the particle $\mathbf{d}^{in}(t)$ we obtain

$$d_x^{in}(\omega) = d_x^{in\,\prime}(\omega)$$

$$d_y^{in}(\omega) = \frac{1}{2}\left(d^{in}{}_y{}'(\omega^+) + d^{in}{}_y{}'(\omega^-) + id^{in}{}_z{}'(\omega^+) - id^{in}{}_z{}'(\omega^-)\right) \qquad (C1)$$

$$d_z^{in}(\omega) = \frac{1}{2}\left(-id^{in}{}_y{}'(\omega^+) + id^{in}{}_y{}'(\omega^-) + d^{in}{}_z{}'(\omega^+) + d^{in}{}_z{}'(\omega^-)\right)$$

Applying the inverse transformation (A2) to the vector $\mathbf{E}^{sp}(t)$ yields

$$E_x^{sp\,\prime}(t) = E^{sp}{}_x(t)$$

$$E_y^{sp'}(t) = E^{sp}{}_y(t)\cos\Omega t + E^{sp}{}_z(t)\sin\Omega t \tag{C2}$$

$$E_z^{sp'}(t) = -E^{sp}{}_y(t)\sin\Omega t + E^{sp}{}_z(t)\cos\Omega t$$

Making use the Fourier transformation in both sides of (C2)

$$\mathbf{E}(\omega) = \int_{-\infty}^{+\infty} dt\,\mathbf{E}(t)\exp(i\omega t) ,$$

we obtain (we omit the argument $\mathbf{r}_0$ of the electric field)

$$E_x^{sp'}(\omega) = E^{sp}{}_x(\omega)$$

$$E_y^{sp'}(\omega) = \frac{1}{2}\left[E^{sp}{}_y(\omega^+) + E^{sp}{}_y(\omega^-) - iE^{sp}{}_z(\omega^+) + iE^{sp}{}_z(\omega^-)\right] \tag{C3}$$

$$E_y^{sp'}(\omega) = \frac{1}{2}\left[iE^{sp}{}_y(\omega^+) - iE^{sp}{}_y(\omega^-) + E^{sp}{}_z(\omega^+) + E^{sp}{}_z(\omega^-)\right]$$

With allowance for the relations in the system $\Sigma'$ related to the particle ($\alpha(\omega)$ is the particle polarizability)

$$d^{in'}{}_x(\omega) = \alpha(\omega)E^{sp'}{}_x(\mathbf{r}_0,\omega)$$
$$d^{in'}{}_y(\omega) = \alpha(\omega)E^{sp'}{}_y(\mathbf{r}_0,\omega) \tag{C4}$$
$$d^{in'}{}_z(\omega) = \alpha(\omega)E^{sp'}{}_z(\mathbf{r}_0,\omega)$$

from (C3), (C4) and (C1) we obtain (when eliminating variables relating to the system $\Sigma'$)

$$d_x^{in}(\omega) = \alpha(\omega)E_x^{sp}(\omega)$$

$$d_y^{in}(\omega) = \frac{1}{2}\left(\alpha(\omega^+)\left[E^{sp}{}_y(\omega) + iE^{sp}{}_z(\omega)\right] + \alpha(\omega^-)\left[E^{sp}{}_y(\omega) - iE^{sp}{}_z(\omega)\right]\right) \tag{C5}$$

$$d_z^{in}(\omega) = \frac{1}{2}\left(-i\alpha(\omega^+)\left[E^{sp}{}_y(\omega) + iE^{sp}{}_z(\omega)\right] + i\alpha(\omega^-)\left[E^{sp}{}_y(\omega) - iE^{sp}{}_z(\omega)\right]\right)$$

Equations (C5) make it possible to use the fluctuation-dissipation theorem for the spontaneous electric field of the surface in the reference system $\Sigma$ of the surface.

**Appendix D: Relation between the Fourier transforms of the fluctuating magnetic vectors in $\Sigma$ and $\Sigma'$**

Applying transformations (A1), (A2) to the fluctuating magnetic vectors $\mathbf{m}^{sp}(t), \mathbf{m}^{in}(t), \mathbf{B}^{sp}(t)$ and making use the same procedures as those made for the vectors $\mathbf{d}^{sp}(t), \mathbf{d}^{in}(t), \mathbf{E}^{sp}(t)$ we obtain

$$m_x^{sp}(\omega) = m_x^{sp\,'}(\omega)$$

$$m_y^{sp}(\omega) = \frac{1}{2}\left(m^{sp}{}_y{}'(\omega^+) + m^{sp}{}_y{}'(\omega^-) + im^{sp}{}_z{}'(\omega^+) - im^{sp}{}_z{}'(\omega^-)\right) \quad \text{(D1)}$$

$$m_z^{sp}(\omega) = \frac{1}{2}\left(-im^{sp}{}_y{}'(\omega^+) + im^{sp}{}_y{}'(\omega^-) + m^{sp}{}_z{}'(\omega^+) + m^{sp}{}_z{}'(\omega^-)\right)$$

$$m_x^{in}(\omega) = \chi(\omega) B_x^{sp}(\omega)$$

$$m_y^{in}(\omega) = \frac{1}{2}\left(\chi(\omega^+)\left[B^{sp}{}_y(\omega) + iB^{sp}{}_z(\omega)\right] + \chi(\omega^-)\left[B^{sp}{}_y(\omega) - iB^{sp}{}_z(\omega)\right]\right) \quad \text{(D2)}$$

$$m_z^{in}(\omega) = \frac{1}{2}\left(-i\chi(\omega^+)\left[B^{sp}{}_y(\omega) + iB^{sp}{}_z(\omega)\right] + i\chi(\omega^-)\left[B^{sp}{}_y(\omega) - iB^{sp}{}_z(\omega)\right]\right)$$

where $\chi(\omega)$ is the magnetic polarizability of the particle.

Relations (B2),(C5),(D1),(D2) and the corresponding fluctuation-dissipation equations, in principle, allows us to calculate all of the values characterizing fluctuation electromagnetic interaction of rotating particles with each another and with different surfaces. In addition, from Eqs. (B2),(C5),(D1),(D2) it follows that $U, \dot{Q}$ and $M_x$ do not contain the cross-terms of electric and magnetic contributions. This makes it possible to calculate the corresponding electric and magnetic terms independently. As far as concerned the magnetic terms, we obtain the expressions which differ from Eqs. (18), (22) and (23) by the replacements $\varepsilon(\omega) \to \mu(\omega), \alpha(\omega) \to \chi(\omega)$, where $\mu(\omega)$ is magnetic polarizability of the surface material.

In the case $\mathbf{\Omega} = (0, 0, \Omega)$, that was considered in our previous work [1,2], the corresponding expressions differ from (B2) and (C5) by the cyclic permutation of indexes $x, y, z$.

FIGURE 1

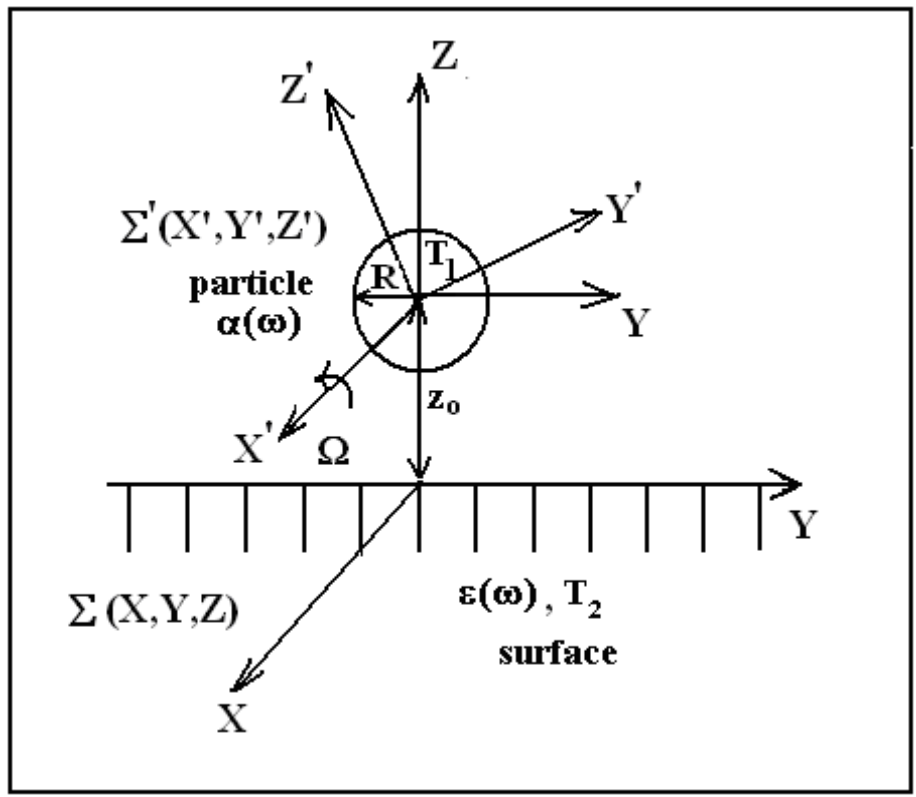

FIGURE 2a

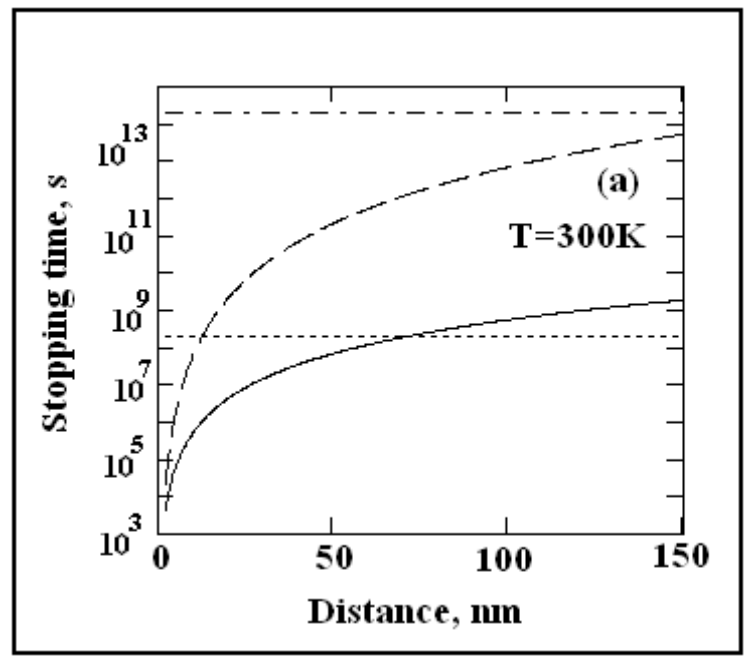

FIGURE 2b

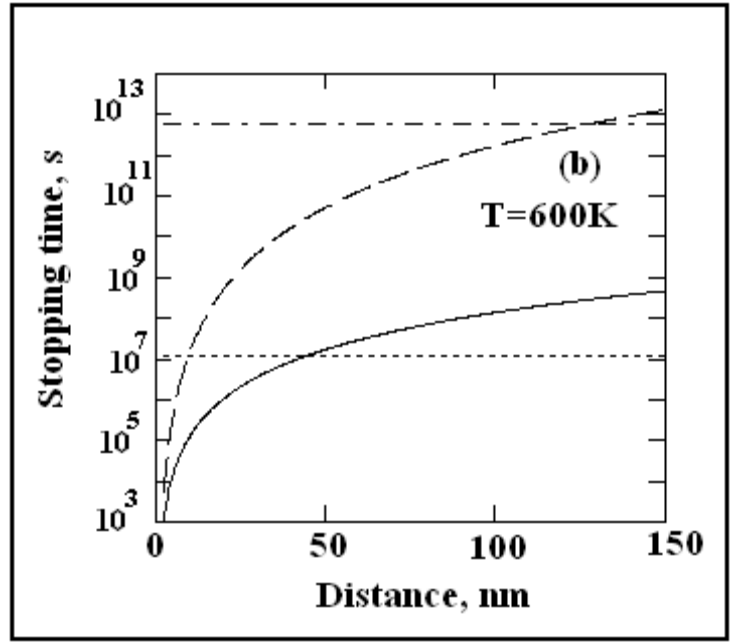

FIGURE 3a

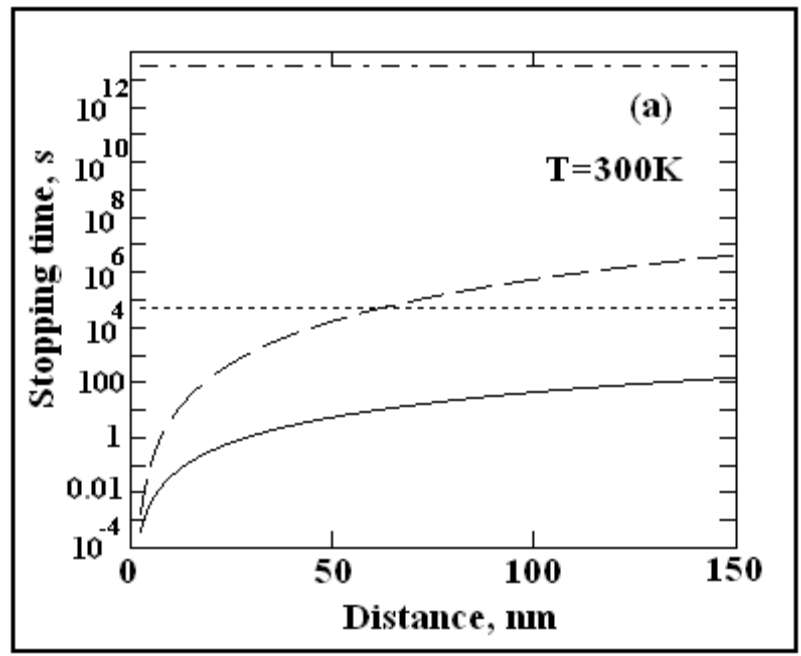

FIGURE 3b

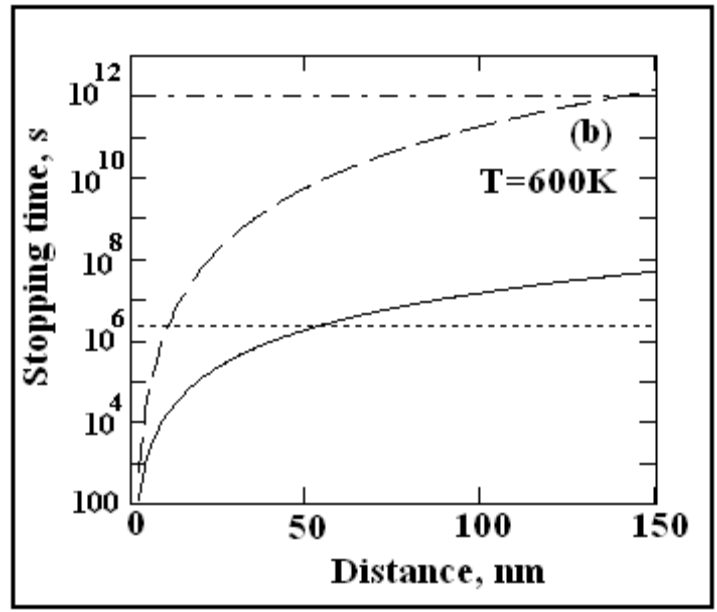

FIGURE CAPTIONS:

Fig. 1 Coordinate systems used and statement of the problem.

Fig. 2 Stopping times of Au particles ($R = 1nm$) near an Ar surface and in vacuum. The solid, dashed, dotted and dash-dotted lines correspond to $\tau_{\Omega,s}$, $\tau_{V,s}$, $\tau_{\Omega,vac}$, and $\tau_{V,vac}$. (a) T=300K; (b) T=600K.

Fig. 3 Same as on Fig.2 for SiC particles near a SiC surface.